\documentclass[12pt,aps,amsmath,latexsym,amsfonts]{JHEP3}

\usepackage{epsfig}
\usepackage{amsfonts,amssymb,amsmath}
\usepackage[hang]{subfigure}
\usepackage{epstopdf}

\newcommand{\be}{\begin{equation}}
\newcommand{\ee}{\end{equation}}
\newcommand{\bea}{\begin{eqnarray}}
\newcommand{\eea}{\end{eqnarray}}

\def\half{\textstyle{\frac{1}{2}}}

\def\quarter{\textstyle{\frac{1}{4}}}

\title{Rotating black hole hair}
\author{
Ruth Gregory$^{1,2}$\thanks{Email: r.a.w.gregory@durham.ac.uk} , 
David Kubiz{\v n}\'ak$^2$\thanks{Email: dkubiznak@perimeterinstitute.ca} , 
Danielle Wills$^1$\thanks{Email: d.e.wills@durham.ac.uk}\\
$^1${\it Centre for Particle Theory, South Road, Durham, DH1 3LE, UK}\\
$^2${\it Perimeter Institute, 31 Caroline Street North, Waterloo, ON, N2L 2Y5,
Canada}
}

\abstract{
A Kerr black hole sporting cosmic string hair is studied in the 
context of the abelian Higgs model vortex. 
It is shown that such a system displays much richer phenomenology 
than its static Schwarzschild or Reissner--Nordstrom cousins, 
for example, the rotation generates a near horizon `electric' field. 
In the case of an extremal rotating black hole, two phases of the Higgs hair 
are possible: Large black holes exhibit standard hair, with the vortex
piercing the event horizon. Small black holes on the other hand, exhibit 
a flux-expelled solution, with the gauge and scalar field remaining 
identically in their false vacuum state on the event horizon. This solution
however is extremely sensitive to confirm numerically, and we conjecture 
that it is unstable due to a supperradiant mechanism similar to the Kerr-adS
instability.
Finally, we compute the gravitational back reaction of the vortex,
which turns out to be far more nuanced than a simple conical deficit.
While the string produces a conical effect, it is conical with respect
to a local co-rotating frame, not with respect to the static frame at
infinity. As a consequence, we find that the ergosphere is shifted,
and geodesics around the black hole are perturbed.
}

\preprint{DCPT-13/05 \\
pi-stronggrv-315}

\begin{document}
\newcommand{\zed}{$\mathbb{Z}_2$}

\section{Introduction}

The ``no hair'' theorems of black holes physics are perhaps one of
the best known examples of a pseudo-theorem, \cite{nohair,nohair2}. 
Although when first proved, 
the conditions placed on the fields seemed reasonable and to cover all
cases of physical interest, it now appears that they were in fact overly
restrictive and there are many cases of physical interest where black holes
can support nontrivial field configurations, and indeed are most stable
doing so. Many applications focus on the case where the black hole remains
asymptotically flat, however, there are two main examples (in 4D)
of interesting non-asymptotically flat hair: the cosmic string and the domain
wall through the black hole \cite{AGK,EGS}. 

Cosmic strings and domain walls  are examples of field theory 
topological defects, solutions to a QFT with a nontrivial vacuum
structure which are topologically stable, hence quasi-classical, 
see, e.g., \cite{VS}.
Each have significant gravitational impact, though not in the
sense of tidal forces: the cosmic string excises a conical 
singularity \cite{NO,Vil81,Garf,AFV,RG}
and the domain wall provides a `mirror' to spacetime, effectively 
compactifying space \cite{IpS,globaldw}. This fact, plus the 
problem of having the fields
essentially end on the event horizon led to the belief that these
objects simply could not enter a black hole or be gravitationally
captured.

The first example of a nontrivial soliton piercing a black hole
was given in \cite{AGK,RGMH}, in which it was shown precisely how the
fields could terminate on the event horizon, and how the back
reaction of the string would give a black hole with a conical deficit
through its poles.
Later works  generalised this to a vortex ending on a black hole,
\cite{EHKT,HR,Roberto,SSS}, (a)dS black holes, \cite{DGM,GBb}, and
to charged black holes, \cite{CCESa,CCESb,BG,BEG,GB},
where a flux expulsion phenomenon was observed for 
extremal Reissner--Nordstrom (RN) black holes of order the string width.
However, at the time the Kerr black hole was not properly explored; not
only was the conventional field ansatz inconsistent in the presence of
rotation, but also the putative conical metric for the back-reacting 
Kerr+vortex system \cite{Aliev} seemed to lead to a singularity of
the vortex energy momentum away from the axis.

Given that most, if not all, of black holes in nature are probably
rotating, this omission is rather glaring! If there is some fundamental
obstruction to a soliton being captured by a black hole if it is rotating,
then this would clearly impact on the properties of cosmic string loops
in a network for example -- which would then have to avoid galaxies
with their central supermassive black holes completely. On the other
hand, if the strings can thread the black hole -- then how do the
core fields accommodate this rotation and its accompanying `electric'
field generation, and is there any analogue of the flux expulsion
of the RN black holes?

In this paper, we show how to correctly thread a vortex through a black
hole: The first technical issue is easily dealt with -- rotation mixes the
time and azimuthal directions near the black hole relative to infinity,
thus the usual angular form of the gauge vector field is coupled
to the zeroth component, and the two cannot be considered
independently\footnote{The paper of Ghezelbash and Mann \cite{GB},
in which charged and/or rotating black holes were considered,
assumed only an angular component of the gauge field and
is thus not a valid ansatz for the rotating black hole.}. 
Indeed, trying to enforce having only an azimuthal
component of the gauge field leads to diverging energy momentum
on the horizon due to a divergent gauge boson norm. We show
how the rotation generates a small electrical flux near the horizon, 
and how the fields respond to increased rotation. We then study the
extremal Kerr limit, exploring whether there is a similar flux
expulsion phenomenon as in the RN black hole. As with RN, we
can demonstrate analytically that there is indeed such a phase transition,
however, a detailed study indicates that unlike RN, this transition appears
to be first order, and the sensitivity of the full numerical system leads us
to suspect that the flux-expelled solution is not dynamically stable, 
and probably has a superradiant instability analogous to 
the Kerr-adS instability found recently \cite{CDLY,CD}.

The second technical issue (of difficulties with the conical deficit
``back-reacted'' metric) is more subtle, and only fully resolved by a
complete and correct back-reaction calculation for the vortex.
Here, we find that the effect of the vortex is {\it not} to cut out a
simple deficit in the $\varphi$ angle, but rather, to alter the length of the
co-rotating azimuthal direction. In essence, to cut out a local co-rotating
deficit angle. Although at first surprising, given the frame dragging
effects of the Kerr metric, this is in fact the most natural outcome
for the string back-reaction. Nonetheless, this leads to surprising and
novel features: The ergosphere of the black hole is shifted, the 
innermost stable circular orbits (ISCO's) are altered, and of course 
any orbit which samples the strong gravity region of the Kerr black hole
is also affected.
Overall, the Kerr
black hole/cosmic string system displays a more interesting and 
rich phenomenology than its Schwarzschild/RN cousins.

\section{Review of the vortex model}

The key feature of a cosmic string is that it is a linear defect, with
energy density equal to tension along its length. Cosmic strings arise
in a range of field theory models, and simply require a nontrivial 
fundamental group of the vacuum manifold, \cite{Vil85}. Cosmic strings also
arise as a relic of brane-antibrane annihilation in brane inflation
models of string theory. While each different symmetry breaking
model or brane inflation model might lead to a different detailed
cosmic vortex, all will have in common this energy/tension balance,
a finite width core of condensate, and some sort of gauge flux 
threading through
(since we require local symmetry breaking for the string to be 
sharply localised). The abelian Higgs model provides a simple and
elegant framework in which to explore cosmic vortices, as it contains
the essential features of the vortex in the simplest possible context.
We therefore use this as our prototype cosmic string.

\subsection{The Nielsen--Olesen vortex}

The Nielsen-Olesen (NO) vortex, \cite{NO}, is the topologically
nontrivial solution of the abelian Higgs model. Its core comprises 
a Higgs condensate threaded with magnetic flux. 
The two cores (scalar and vector) 
in general have different widths, given by the inverse Higgs and gauge 
boson masses, and the ratio determines whether the vortex 
is type I, II, or supersymmetric (Bogomolnyi limit, \cite{Bog}).

The abelian Higgs action is\footnote{We use units in which $\hbar=c=1$ and a
mostly minus signature.}
\be \label{abhact}
S = \int d^4x \sqrt{-g} \left [ D_{\mu}\Phi ^{\dagger}D^{\mu}\Phi -
{\quarter} {\tilde F}_{\mu \nu}{\tilde F}^{\mu \nu} - {\quarter}\lambda
(\Phi ^{\dagger} \Phi - \eta ^2)^2 \right ]\,,
\ee
where $\Phi$ is the Higgs field, and $A_\mu$ the U(1) gauge 
boson with field strength ${\tilde F}_{\mu\nu}$.
As per usual, we rewrite the field content as:
\bea
\Phi (x^{\alpha}) &=& \eta  X (x^{\alpha}) e^{i\chi(x^{\alpha}) }\,,  \\
A_{\mu} (x^{\alpha}) &=& \frac{1}{e} \bigl [ P_{\mu} (x^{\alpha}) - 
\nabla_{\mu} \chi (x^{\alpha}) \bigr ]\,.
\eea
These fields extract the physical degrees of freedom of the broken symmetric
phase, with $X$ being the residual massive Higgs field, and $P_\mu$ the 
massive vector boson.
$\chi$, as the gauge degree of freedom is explicitly subtracted,
although any non-integrable phase factors have a physical interpretation as
a vortex.

In terms of these new variables, the equations of motion are
\bea \label{vorteqn}
\nabla _{\mu}\nabla ^{\mu} X - P_{\mu}P^{\mu}X + \frac{\lambda\eta^2}{2}
X(X^2 -1) &=& 0\,, \\
\nabla _{\mu}F^{\mu \nu} + \frac{X^2 P^{\nu}}{\beta} &=& 0\,,
\eea
where $\beta = \lambda/2e^2$ is the Bogomol'nyi parameter~\cite{Bog}, and
$F_{\mu\nu}$ is the field strength of $P_\mu$.

The Nielsen--Olesen vortex is a (flat space) solution to these equations 
expressed in cylindrical polar coordinates as:
\be
X = X_0(R)\,, \qquad P_\mu = P_0(R)\partial_\mu\phi\,,
\label{xpform}
\ee
where $R=r\sqrt{\lambda}\eta$, and
$X_0$ and $P_0$ satisfy
\be \label{basic}
\begin{aligned}
-X''_0 - \frac{X'_0}{R} + \frac{X_0P^2_0}{R^2} 
+ {\half} X_0(X^2_0-1) &=& 0\,, \\
-P''_0+\frac{P'_0}{R} + \frac{X^2_0P_0}{\beta} &=& 0\,. 
\end{aligned}
\ee
See figure \ref{fig:NOV} for a plot of $X_0$ and $P_0$ for $\beta=1$. 
\FIGURE{
\includegraphics[scale=0.5]{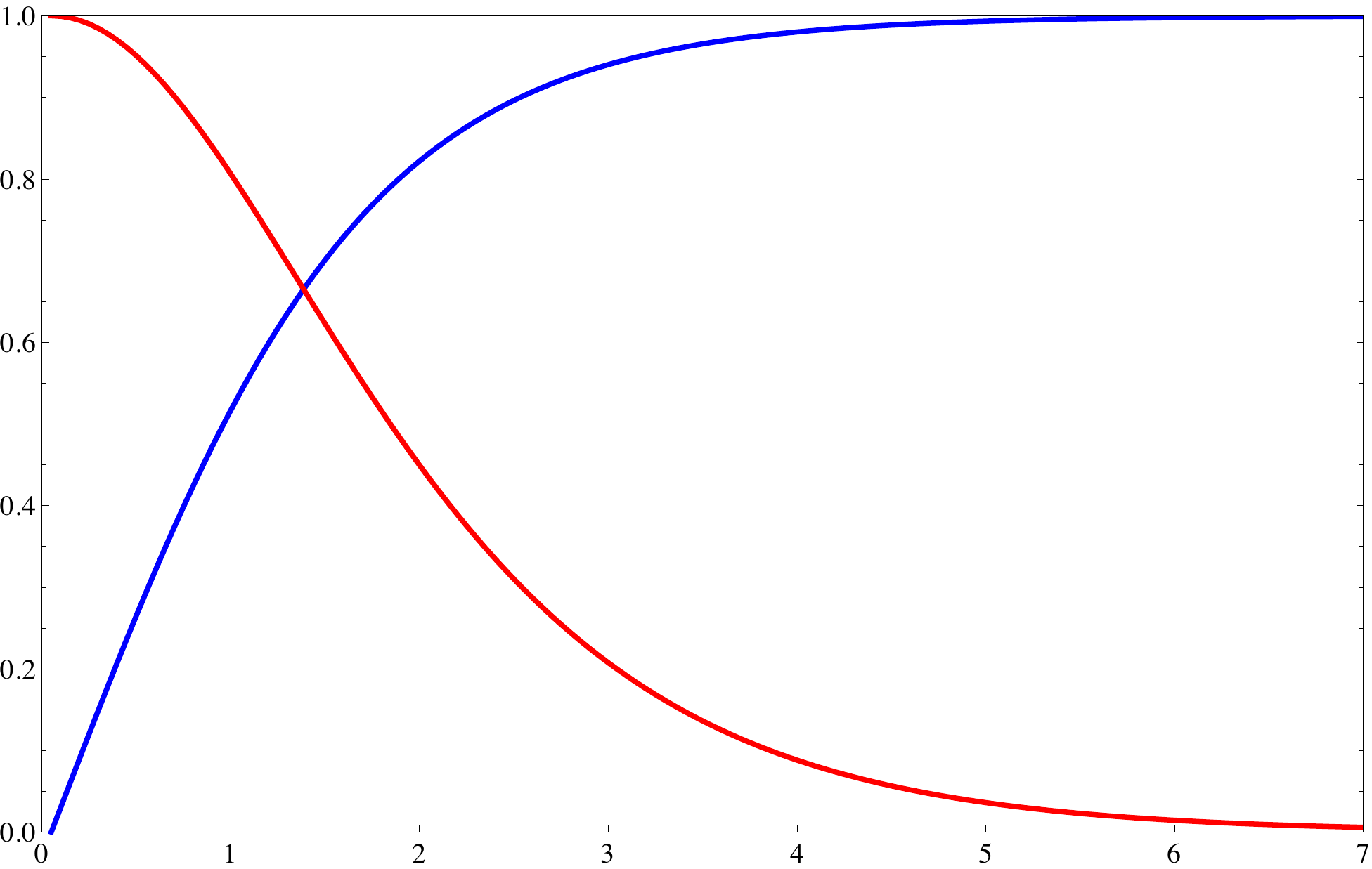}
\caption{Numerical solution of the Nielsen-Olesen vortex: $X_0$ (blue) and
$P_0$ (red). Notice $P_0$ falls off a little more slowly on this scale.}
\label{fig:NOV}
}

For later convenience, we give a lightning (but useful)
review of the gravitational effect of this vortex.
The idea here is to solve the Einstein equations,
\be
R_{\mu\nu} - \frac12 R g_{\mu\nu} = 8\pi G T_{\mu\nu}\,,
\ee
simultaneously to the curved space abelian-Higgs vortex equations.
The gravitational effect of the vortex is determined by the 
dimensionless ratio
\be
\epsilon = 8\pi G\eta^2\,,
\label{epsilondef}
\ee
which will typically be of order $10^{-7} - 10^{-12}$ for cosmic
strings of cosmological relevance. Thus, we can perform an expansion
in $\epsilon$, finding the background (flat space) Nielsen-Olesen
solution, and using its energy momentum to compute the leading order
gravitational correction to flat space.

Looking for a static solution, we can choose a gauge in which 
the metric takes the form:
\be
ds^2 = e^{2\lambda} dt^2 - e^{2(\nu-\lambda)} [ dz^2+dR^2 ]
- \alpha^2 e^{-2\lambda} d\phi^2\,,
\label{selfgravG}
\ee
with curvature:
\bea
\sqrt{g} (R^\phi_\phi + R^t_t ) &=& \alpha ''\,, \\
\sqrt{g} R^t_t &=& [ \alpha \lambda ']'\,, \\
\sqrt{g} R^z_z &=& [ \alpha (\nu -\lambda) ']'\,, \\
\sqrt{g} R^R_R &=& \alpha''+ \alpha (\nu'' -\lambda'') 
-\alpha'(\lambda'+\nu') + 2 \alpha \lambda '^2\,, 
\eea
where $\alpha=R$, $\lambda = \nu = 0$ to leading order.

The energy-momentum tensor of the vortex can readily be computed 
to leading order as:
\be
\begin{aligned}
T^t_t = T^z_z = {\cal E} &=  X'^2 + \frac{X^2 P^2}{R^2} 
+ \beta \frac{P'^2}{R^2} + \frac14 (X^2-1)^2\,, \\
T^R_R = -{\cal P}_R &=  -X'^2 + \frac{X^2 P^2}{R^2} - \beta \frac{P'^2}{R^2} 
+ \frac14 (X^2-1)^2\,, \\
T^\phi_\phi = -{\cal P}_\phi &=  X'^2 - \frac{X^2 P^2}{R^2} 
- \beta \frac{P'^2}{R^2} + \frac14 (X^2-1)^2\,, 
\end{aligned}
\label{NOTab}
\ee
and a useful identity from the equations of motion \eqref{basic} is
\be
\frac{d~}{dR} [ R{\cal P}_R ] = {\cal P}_\phi\,.
\label{PRPphi}
\ee
Solving the Einstein equations to leading order with this energy momentum
tensor is then straightforward, and gives
\bea
\alpha &=& \left [ 1 - \epsilon \int_0^R R({\cal E} - {\cal P}_R) dR\right]R
+\epsilon \int_0^R R^2 ({\cal E}- {\cal P}_R)dR\,, \label{SGalpha}\\
2\lambda &=& \epsilon \int_0^R R{\cal P}_R dR = \nu\,. \label{SGlambda}
\eea
It is then easy to see the conical nature of this spacetime, as the
exponential fall off of the $X$ and $P$ fields mean that the integrals
converge rapidly, and the asymptotic form of the metric is
\be
ds^2 = d{\hat t}^2 - d{\hat z}^2 - d{\hat R}^2 - {\hat R}^2 
(1-\epsilon {\hat \mu})^2 d\phi^2\,,
\ee
where the coordinates have been rescaled (${\hat t}
= e^{\epsilon \lambda_\infty} t$ etc.) to those of an asymptotic observer,
and 
\be
{\hat \mu} = 1 - \frac{d~}{dR} \alpha_\infty + 2\lambda_\infty
= \int_0^\infty R\, {\cal E} dR
\label{muhat}
\ee
is the renormalised energy per unit length of the cosmic string\footnote{The
actual energy per unit length is $\eta^2{\hat\mu}$.}.
Note that the effect of the transverse stresses of the string is to 
alter the details of the metric response, but that these details
cancel out to leave the headline result that the conical deficit 
depends only on the energy per unit length of the string. 
For the Bogomolnyi limit $\beta=1={\hat\mu}$, these stresses vanish, 
and the string geometry is flat in the parallel ($t,z$) directions, 
and a smooth snub-nosed cone in the transverse ($R,\phi$) directions.

\subsection{Cosmic string and Schwarzschild black hole}

The basic idea of putting the vortex on the black hole is to first
find an approximate solution assuming the string is much thinner
than the horizon. There are (strictly speaking)
three length scales, the two string core widths as already
mentioned, and the black hole scale, however, by fixing the Bogomolnyi 
parameter and setting our scale to the string width, only one dimensionless
parameter remains relevant: the black hole horizon radius, $r_+$, 
in units of string width (or $2GM\sqrt{\lambda}\eta$).

One then writes the vortex equations in the background of the 
Schwarzschild black hole:
\bea
-\Bigl( 1 \!-\! \frac{r_+}{r} \Bigr)
X_{,rr} -\frac{2r\!-\!r_+}{r^2}X_{,r}
- \frac{X_{,\theta\theta}}{r^2} - \frac{\cot\theta X_{,\theta}}{r^2}
\!+\! \frac12 X(X^2 \!-\!1) 
+  \frac{XP^2_\phi}{r^2\sin^2\!\theta} 
&=&0\,,\qquad\\ 
\left ( 1 \!-\! \frac{r_+}{r} \right ) \partial_r\partial_r P_{\phi}
+\frac{1}{r^2}\partial_\theta \partial_\theta P_\phi
+\frac{r_+}{r^2}\partial_r P_\phi 
-\frac{\cot \theta}{r^2}
\partial_\theta P_\phi-\frac{X^2P_\phi}{\beta} 
&=&0\,, \qquad\ \ 
\eea
and solves numerically. As noted in \cite{AGK}, for large $r_+$, these 
equations have a very good approximate solution of the form\footnote{%
In fact, the solution is valid as long as $\frac{r_+}{r}\sin^2\!\theta \ll 1$. 
Therefore, even for small $r_+$ it is valid close to the poles, 
or sufficiently far away from the black hole. 
} 
\be\label{Schwappro}
X\simeq X_0(r\sin\theta)\,,\quad 
P_\phi\simeq P_0(r\sin\theta)\,. 
\ee
Both this approximate solution, and the numerical integration, show
that the vortex core is surprisingly undisturbed by the black hole,
and the flux lines appear to simply ``go through'' the black hole 
(see figure~\ref{fig:SCH}).
\FIGURE{
\includegraphics[scale=0.5]{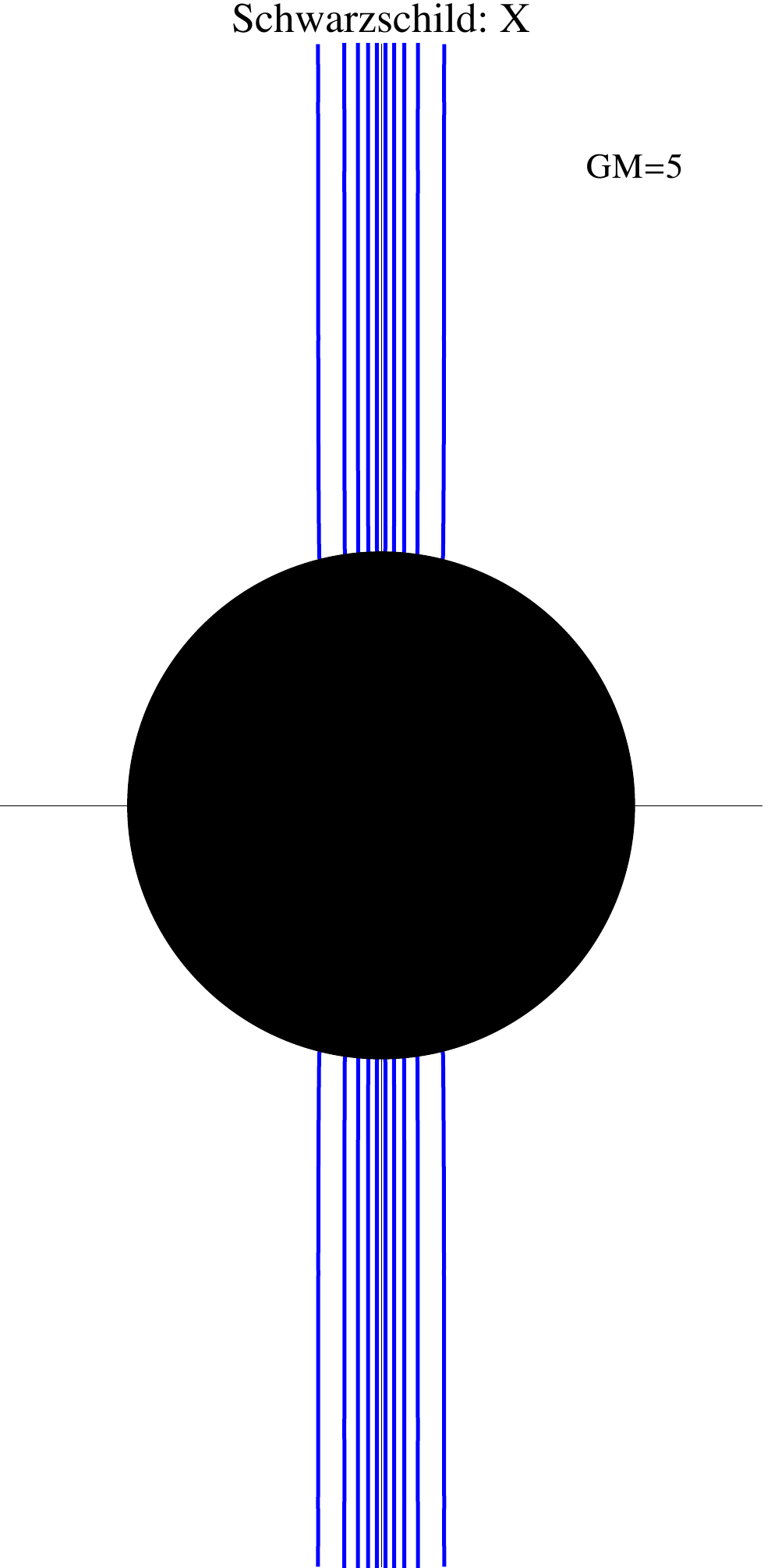}\hskip 2cm
\includegraphics[scale=0.5]{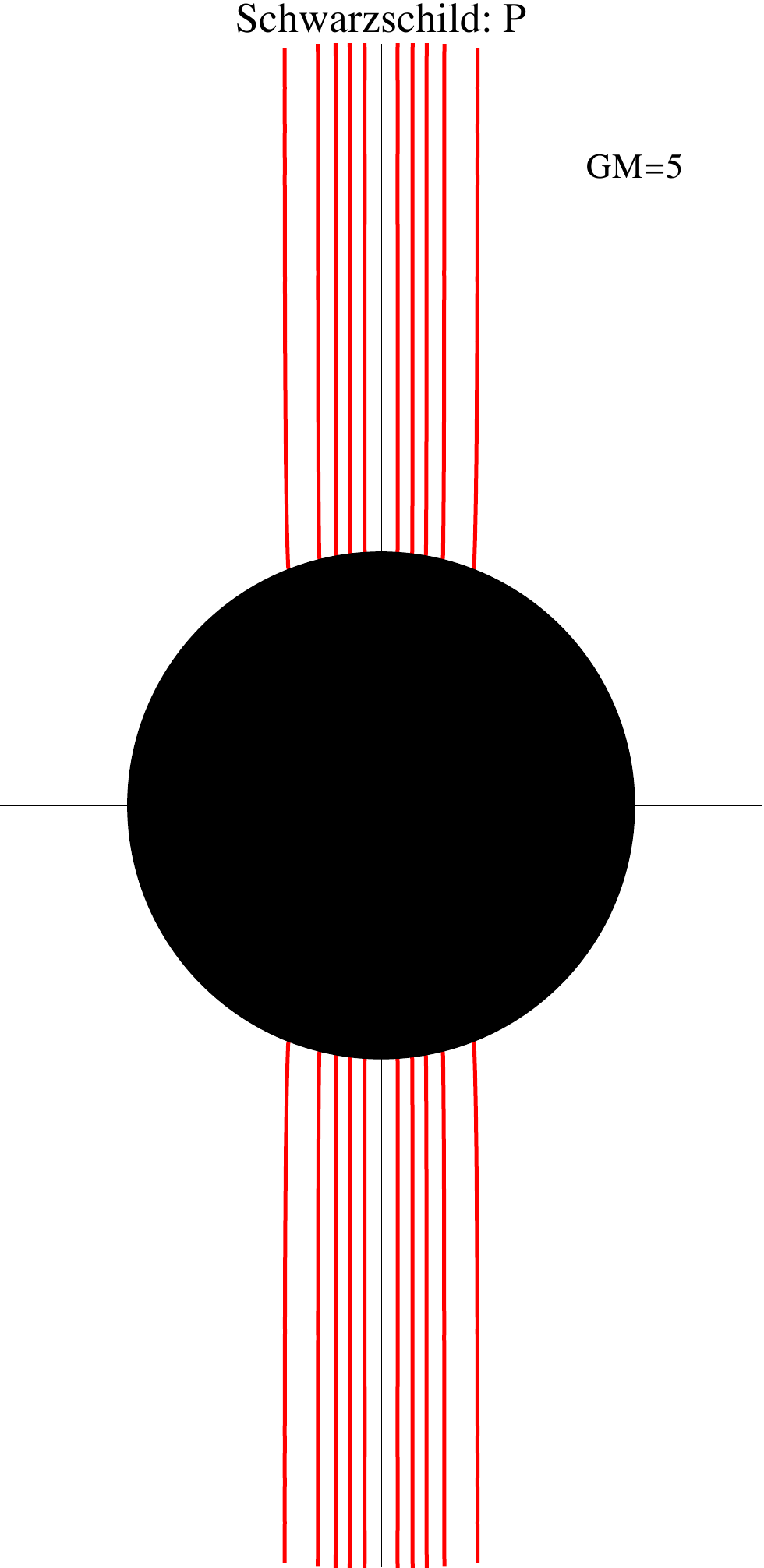}
\caption{%
The equipotentials of the NO vortex in the Schwarzschild background.
The Higgs contours are in blue, and the $P_\phi$ contours are in red. 
In each case contours are shown for $X,P_\phi =0.1, 0.3, 0.5, 0.7, 0.9$. 
} \label{fig:SCH}  
}

\subsection{Flux expulsion: extremal RN black hole}\label{sec:RN} 

When a small electric charge is added to the black hole the metric becomes
the Reissner-Nordstrom (RN) solution, and there is no 
qualitative difference in how the string pierces the black hole, with ansatz 
\eqref{Schwappro} remaining a very good approximation to the exact solution 
for large RN black holes. However, when the black hole becomes {\em extremal} 
a new interesting phenomenon occurs: whereas for large extremal RN black 
holes the string still threads the horizon, below a certain critical mass,
or black hole radius $r_c$, both the Higgs and the $U(1)$ fields are 
{\em expelled} from the black hole.    
The reason for this behavior is that in the extremal case, the 
horizon equations actually {\em decouple} from the exterior geometry \cite{BEG} and admit a flux expulsion solution.
In fact, the authors of \cite{BEG} were able to place analytic bounds 
and demonstrate that for $r_+<0.7$ the expulsion must occur whereas 
for $r_+>2.9$ the penetration is inevitable. Numerical work actually 
places this threshold at about $r_c\approx 1.9$.

Since the discussion of this interesting behaviour is in some sense 
analogous to what we shall see in the extremal Kerr case let us 
recapitulate some of the features of this calculation.
The vortex field equations in the RN background read
\bea\label{RN}
-\frac{1}{r^2}\partial_r ( \Delta X_{,r})
-\frac{1}{r^2\sin\theta}\partial_\theta(\sin\theta X_{,\theta})
+\frac{XP_\phi^2}{r^2\sin^2\!\theta}-\frac{1}{2}X(1-X^2)&=&0\,,\\
\partial_r\left(\frac{\Delta}{r^2}\partial_r P_\phi\right)
+\frac{\sin\theta}{r^2}\partial_\theta \left(\frac{\partial_\theta P_\phi}
{\sin\theta}\right)-\frac{X^2P_\phi}{\beta}&=&0\,,
\eea
where $\Delta = r^2 - 2GMr + GQ^2=r^2 g_{tt}$.
Expanding near the horizon in the extremal case, when the metric 
function $\Delta$ has a double root $\Delta=(r-r_+)^2$,
\be
X=\xi_0(\theta)+(r-r_+)\xi_1(\theta)+\dots\,,\quad 
P_\phi=\pi_0(\theta)+(r-r_+)\pi_1(\theta)+\dots\,,
\ee
the horizon equations decouple from the exterior geometry, 
giving\footnote{The existence of a double root of function $\Delta$ 
is crucial for such a decoupling.} 
\be
\label{RNhorizon}
\begin{aligned}
\xi_0''+\cot\theta\xi_0'-\frac{\xi_0\pi_0^2}{\sin^2\!\theta}
+\frac{r_+^2}{2}\xi_0(1-\xi_0^2)&=0\,,\\
\pi_0''-\cot\theta \pi_0'-\frac{r_+^2}{\beta}\xi_0^2\pi_0&=0\,,
\end{aligned}
\ee
where $\xi_0$ and $\pi_0$ must be symmetric around $\theta=\pi/2$, 
and obey $\xi_0=0, \pi_0=1$ at $\theta=0, \pi$.
Obviously, such equations admit the flux expulsion solution $\xi_0=0, 
\pi_0=1$ everywhere. However, such a solution must extend to the bulk, 
which, as we shall see, is possible only for $r_+<r_c$. 

To see this, suppose that expulsion occurs, i.e.\ on the horizon $X\equiv 0$,
$P_\phi\equiv 1$, with $X$ increasing and $P_\phi$ decreasing towards their
asymptotic values away from the horizon. Now consider a region close
to the horizon in which $\partial_r ( \Delta X_{,r})>0$  and $X^2\ll 1$, 
then from \eqref{RN} 
\be\label{cond}
X>XP^2_\phi>\sin\theta\partial_\theta(\sin\theta X_{,\theta})
+\frac{1}{2}{r_+^2}\sin^2\!\theta X\,.
\ee
Since $\sin\theta X_{,\theta}=0$ at $\theta=0, \pi/2$, and is positive for
small $\theta$, its derivative must have at least one zero on $(0,\pi/2)$, so
define $\theta_0<\pi/2$ as the first value of $\theta$ at 
which $\partial_\theta(\sin\theta X_{,\theta})=0$. From (\ref{cond}),
$\frac{1}{2}r_+^2\sin^2\!\theta_0<1$, which is manifestly true for 
$r_+<\sqrt{2}$ so let us consider a larger black hole with $r_+>\sqrt{2}$, and 
define $\alpha>\theta_0$ by $r_+^2\sin^2\!\alpha=2$. Then, integrating 
\eqref{cond} on the range $(\theta, \pi/2)$, for $\theta>\alpha$ gives 
\be
X_{,\theta}>\frac{1}{\sin\theta}\int_\theta^{\pi/2}\!\!\!\!
d\theta' X(\theta')\left(\frac{r_+^2}{2}\sin\theta
-\frac{1}{\sin\theta}\right)>
\frac{X(\theta)}{\sin\theta}\int_\theta^{\pi/2}\!\!\!\!
d\theta'\left(\frac{r_+^2}{2}\sin\theta-\frac{1}{\sin\theta}
\right)\,,\nonumber
\ee
i.e., 
\be
X_{,\theta}(\theta)>X(\theta)
\left[\frac{r_+^2}{2}\cot\theta+\frac{\ln\tan(\theta/2)}{\sin\theta}
\right]\,.
\ee
Due to the fact that $X_{,\theta\theta}<0$ on $[\theta_0, \pi/2]$ 
and $X_{,\theta}(\theta)<\frac{X(\theta)-X(\alpha)}{\theta-\alpha}
<\frac{X(\theta)}{\theta-\alpha}$, for consistency we must have
\be
1>(\theta-\alpha)\left[\frac{\cot\theta}{\sin^2\!\alpha}
+\frac{\ln\tan(\theta/2)}{\sin\theta}\right]
\ee
over the range $\theta\in(\alpha,\pi/2)$. One finds this is violated 
for $r_+^2>8.5$. Hence for $r_+\geq \sqrt{8.5}\approx 2.92$ 
the vortex must pierce the horizon.

A lower bound for $r_c$ can be obtained by considering the horizon 
equations \eqref{RNhorizon}. Namely, let a piercing solution of these 
equations exist. The second equation implies that $\pi_0$ monotonically 
decreases and reaches its first minimum $\pi_m\geq 0$ at $\theta=\pi/2$.
Let us further assume that $\xi_0$ monotonically increases and reaches its 
first maximum $\xi_M\leq 1$ at $\theta=\pi/2$.\footnote{%
This assumption seems plausible based on energetic considerations: 
if the scalar field produced some ``wobbles'', having for example a 
first maximum for $\theta<\pi/2$ and then went to a minimum at $\pi/2$, 
we expect this to be less energetically favorable.
}  Then one can derive, \cite{BEG}, that $r_c^5/(\sqrt{2}-r_c)^2 \geq 
\beta^2/\sqrt{2}$, giving $r_c  \simeq 0.7$ for $\beta=1$. 

Numerical work shows that (taking $\beta=1$) a transition between the penetration and 
expulsion actually occurs for $r_c\approx 1.9$, in which 
case $\pi_m\approx 1$ and $\xi_M\approx 0$. Such a transition is 
therefore continuous from the point of view of the fields on 
the horizon. The RN flux expulsion phase transition is indicated
in figure \ref{fig:RNK}, where it is compared to the Kerr case. 
It is worth remarking on the response of this phase transition to the
Bogomolnyi parameter, $\beta$. 
As $\beta$ drops, the gauge core becomes more confined, and thus we see a
drop in the critical mass before the black hole becomes small enough to sit
inside the magnetic flux core.
On the other hand, for $\beta\geq{\cal O}(1)$, the gauge core 
is more diffuse, leading to a behaviour in the Higgs field more 
analogous to a global vortex.
The order parameter (the value of the Higgs field at $\theta=\pi/2$)
drops more smoothly, before finally the flux expulsion kicks in
when the black hole finally comes within the Higgs core, at roughly
the same critical mass as for a $\beta=1$ vortex.
We shall see that all these features (continuity, $\beta$-dependence) 
are substantially different for the extremal Kerr black hole.

\section{Higgs hair for the Kerr black hole}

\subsection{Approximate solution}

The Kerr geometry [in $(+,-,-,-)$ signature] reads
\be
ds^2 = \frac{\Delta\!-\!a^2\sin^2\!\theta}{\Sigma}dt^2
+\frac{4GMar\sin^2\!\theta}{\Sigma} dtd\varphi\!-\!\Sigma d\theta^2
-\frac{\Gamma}{\Sigma}\sin^2\!\theta \,d\varphi^2\!
-\!\frac{\Sigma}{\Delta}dr^2\,,\quad\ 
\ee
where $a=J/M$ and 
\be
\Sigma=r^2+a^2\cos^2\!\theta\,,\quad \Delta=r^2-2GMr+a^2\,\qquad
\Gamma = (r^2+a^2)^2 - \Delta a^2 \sin^2\theta \,.
\ee
Due to the rotation, we expect a mixing between the $t$ and $\phi$ degrees
of freedom, so we consider both a nonzero $P_\phi$ {\it and} $P_t$:
\bea
\frac{X^2}{\beta} P_\phi&=&
\frac{\Delta}{\Sigma}\partial_r \partial_r P_\phi 
+\frac{1}{\Sigma}\partial_\theta \partial_\theta P_\phi
+\frac{2GM\rho^2}{\Sigma^3} (r^2 - a^2 \cos^2\!\theta)\partial_r P_\phi\nonumber\\
&&-\frac{\cot \theta}{\Sigma^3} 
(\Sigma^2 + 4GMra^2\sin^2\!\theta)\partial_\theta P_\phi 
-\frac{4a^3GMr}{\Sigma^3}\cos\theta \sin^3\!\theta
\partial_\theta P_t\\
&&+\frac{2GMa\sin^2\!\theta}{\Sigma^3}\bigl[2 r^2 \Sigma 
+ \rho^2(r^2 - a^2\cos^2\!\theta )\bigr]\partial_r P_t\,,\nonumber\\
%%%%%
\frac{X^2}{\beta}P_t&=&
\frac{\triangle}{\Sigma}\partial_r \partial_r P_t 
+\frac{1}{\Sigma}\partial_\theta \partial_\theta P_t
+ \frac{4GMra}{\Sigma^3}\cot\theta \left (\partial_\theta P_\phi 
+ a \sin^2\!\theta \partial_\theta P_t \right )
+\frac{\cot\theta}{\Sigma} \partial_\theta P_t\nonumber\\ 
&&+\frac{2GMa}{\Sigma^3}(\Sigma - 2r^2)\partial_r P_\phi-
\frac{1}{\Sigma^3}\bigl[2GM(2r^2\rho^2-a^2\sin^2\!\theta\Sigma)
-2r\Sigma^2\bigr]\partial_r P_t \,,\label{P0eq}\qquad\qquad\\
%%%%%
0&=&\frac{\Delta}{\Sigma} X_{,rr} + \frac{2(r-GM)}{\Sigma} X_{,r}
+\frac{X_{,\theta\theta}}{\Sigma} + \frac{\cot\theta X_{,\theta}}{\Sigma}
+ \frac12 X(1-X^2) + X P_\mu^2\,,\qquad \label{Xeq}
\eea
where $\rho^2=r^2+a^2$ has been introduced for visual clarity, and
\be\label{psquared}
P_\mu^2 = \frac{(\rho^2P_t + aP_\phi)^2}{\Sigma\Delta}
- \frac{(P_\phi+a \sin^2\!\theta P_t)^2}{\Sigma \sin^2\!\theta}\,.
\ee
We now see explicitly why we needed to introduce the $P_t$ field
(indeed, this was first noted by Wald \cite{Wald} who found an expression 
for constant probe magnetic flux field through a Kerr black hole).
Clearly (\ref{P0eq}) does not allow $P_t=0$ unless $a=0$. 
Indeed, a little investigation shows that the approximate 
analytic solution (\ref{Schwappro}) can be generalized in the Kerr case to
\be\label{ApproxSol}
X \simeq X_0(R)\,,\quad  
P_\phi \simeq P_0(R)\,,\quad  
P_t \simeq -\frac{2GMar}{\rho^4} P_0(R)\,,\quad R\equiv \rho\sin\theta.
\ee
Figure~\ref{fig:app2num} illustrates how good an approximation
to the full numerical solution this expression is.
\FIGURE{
\includegraphics[scale=0.75]{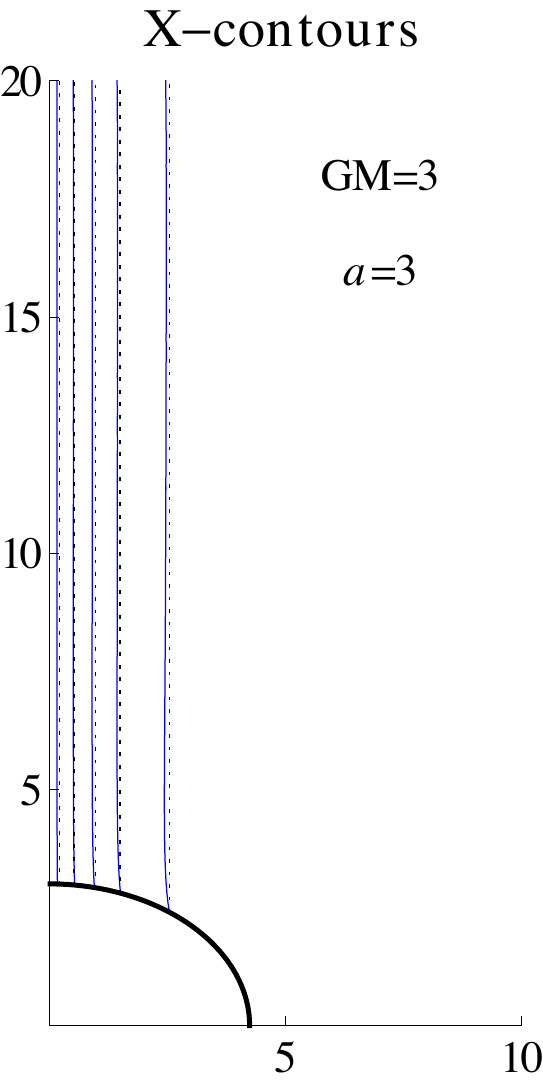}
\includegraphics[scale=0.75]{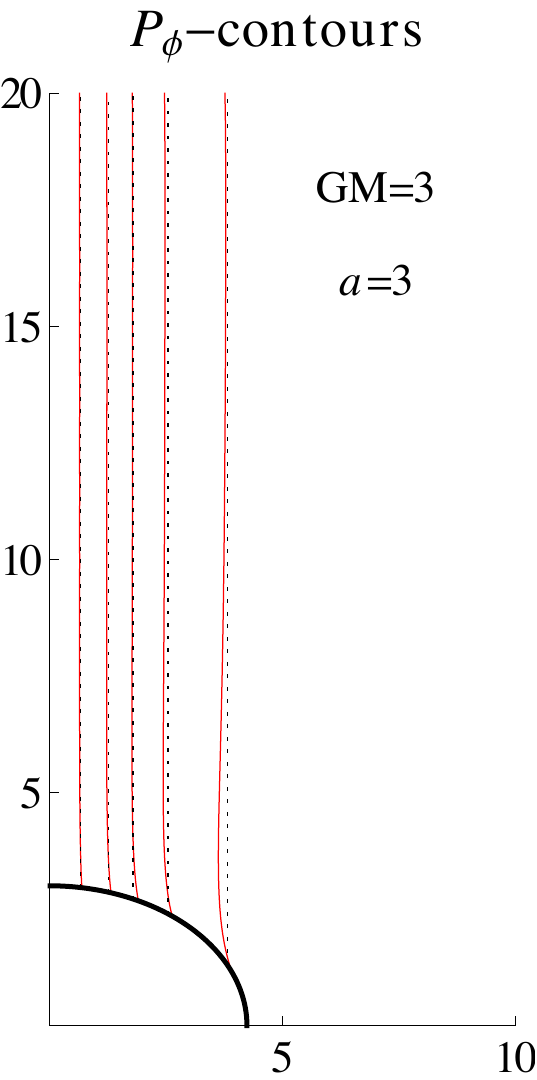}
\includegraphics[scale=0.75]{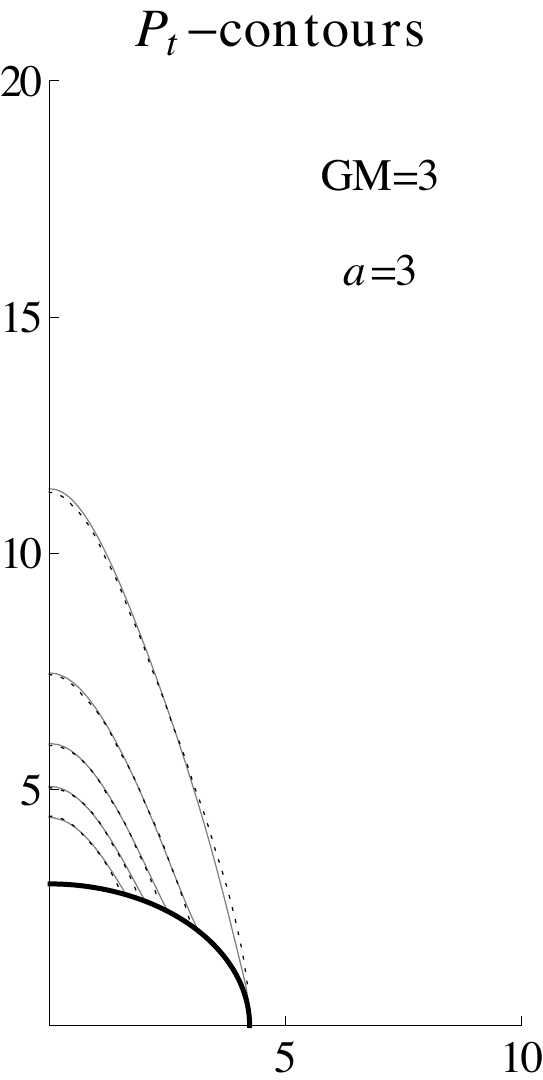}
\caption{
A comparison of the approximate and exact numerical solutions
for an extremal $GM=a=3$ Kerr black hole. In spite of the low
value of black hole mass, \eqref{ApproxSol} is still an
extremely good approximation to the actual result. Here,
the Higgs contours are in blue, the $P_\varphi$ contours in red,
the $P_t$ contours in grey, and all the corresponding approximate
solution contours in dashed black. Contours are shown for $X,P_\varphi
=0.1, 0.3, 0.5, 0.7, 0.9$, and for $P_t = -0.099, -0.077,
-0.055, -0.033, -0.011$.} 
\label{fig:app2num}  
}

\subsection{Numerical solution}

In order to demonstrate conclusively that the abelian
Higgs vortex is compatible with the rotating black hole, we need
to numerically integrate the equations of motion. Since this is an
elliptic problem we used a gradient flow method on a polar
grid, updating the event horizon as per the method of \cite{AGK}
with the constraint that on the horizon 
\be
P_t=-\frac{aP_\phi}{r_+^2+a^2}\,.
\ee
For the Kerr black hole however, there is also an additional subtlety:
the vortex boundary conditions ($X=0$, $P_\varphi=1$) placed on axis 
only restrict the $X$ and $P_\varphi$ fields, and not $P_t$.
This is not surprising and represents the fact that there is a
dyonic degree of freedom the black hole introduces to the solution.
(This also exists in the Schwarzschild set-up, but was not noticed as 
the electric and magnetic degrees of freedom of the gauge boson
decouple there.) Since we do not wish to pick up a spurious charge
of the black hole, we allow the $P_t$ field to relax freely, and update
it along the axis by continuity, thus ensuring that $P_t$ is only as 
big as it needs to be to counter the magnetic part of the vortex.
Figure \ref{fig:XPm5} shows some sample solutions for a large-ish
black hole both at, and away from, extremality.
\FIGURE{
~~\includegraphics[scale=0.8]{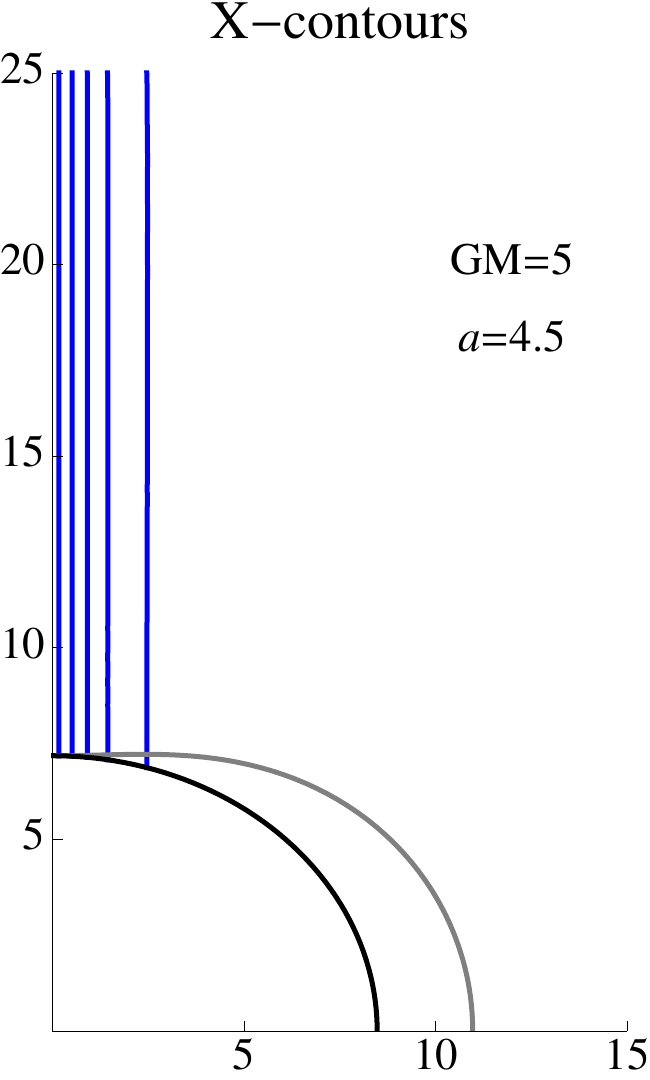}~~
\includegraphics[scale=0.8]{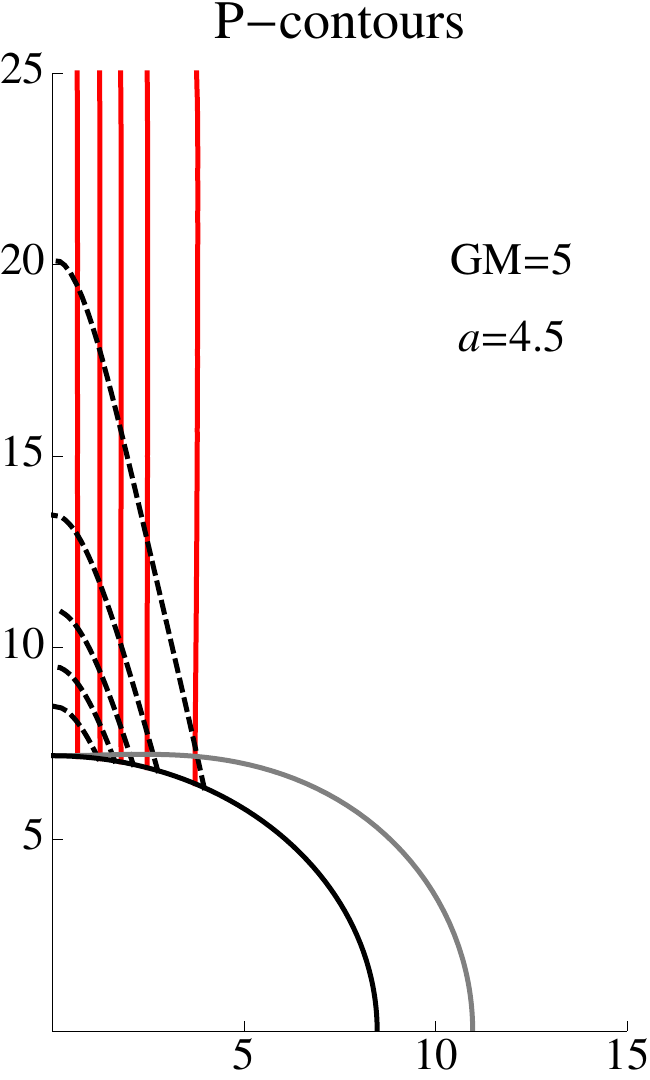}\\
\includegraphics[scale=0.75]{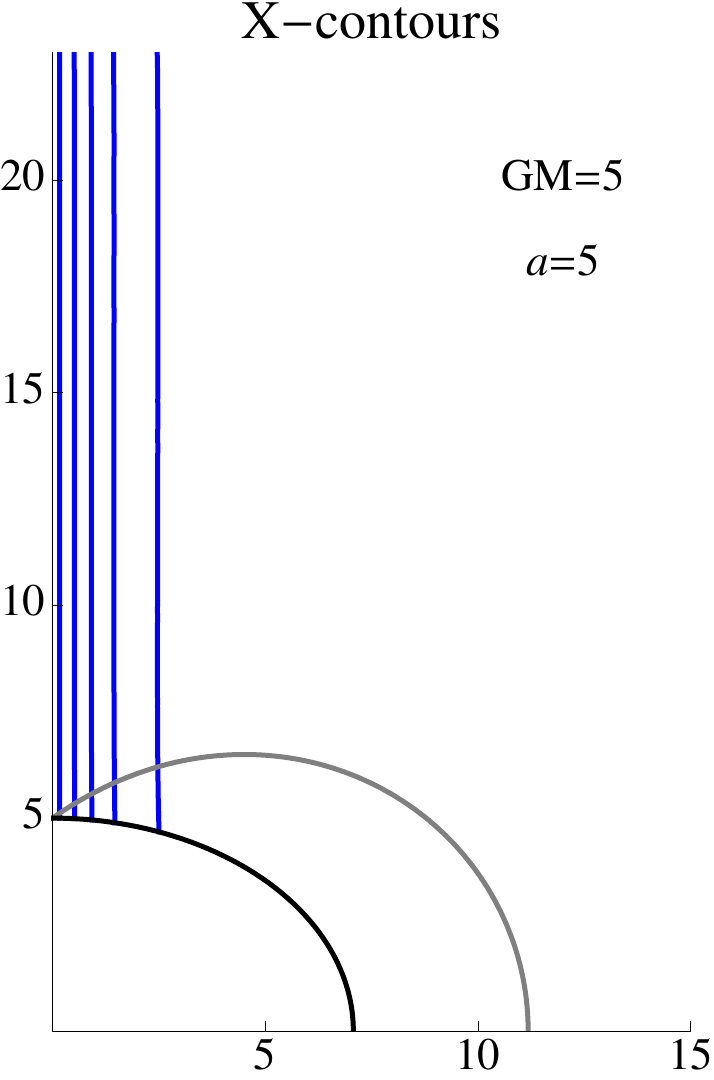}~~
\includegraphics[scale=0.75]{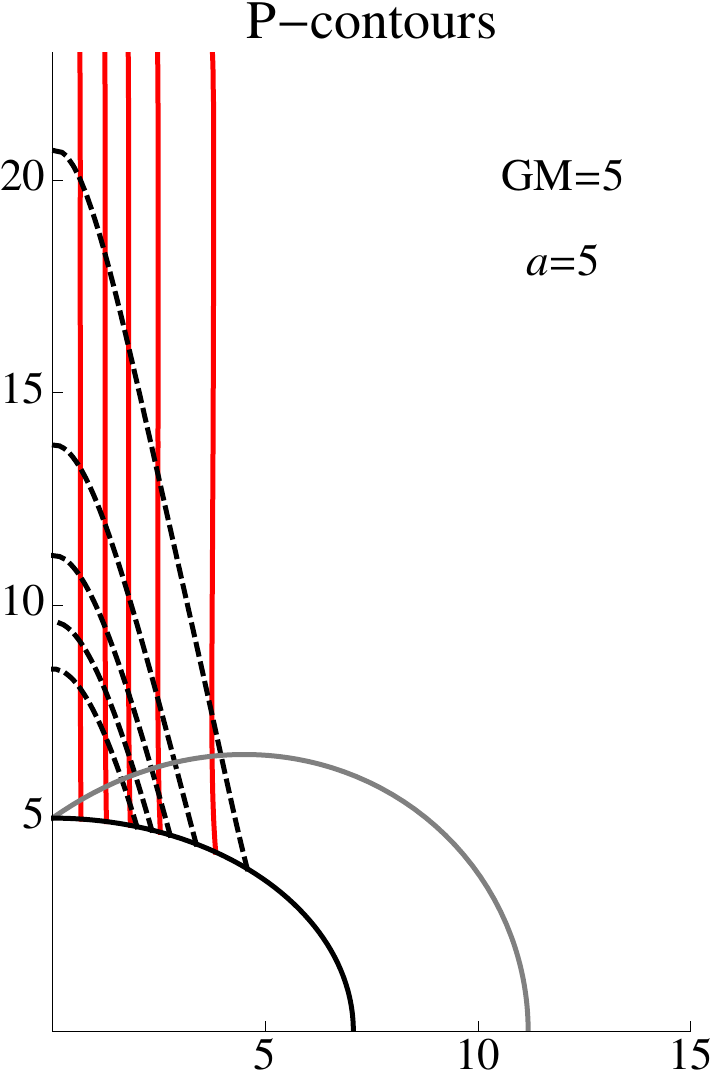}
\caption{Numerical solution for a Kerr black hole with the
values of $GM$ and $a$ indicated. On the left, 
the $X=0.1,0.3,0.5,0.7,0.9$ contours are plotted in blue. On the
right, the $P_\varphi=0.1,0.3,0.5,0.7,0.9$ contours are in red, and the
$P_t=-0.045,-0.035,-0.025,-0.015,-0.005$ contours are in dashed black.
The horizon is shown in black, and the edge of the ergosphere in grey.}
\label{fig:XPm5}
}

\section{Extremal Kerr black holes}

Having shown that the vortex can sit through a black hole, at the
price of some induced electric field, it is interesting to look at the
extremal limit of the Kerr black hole in more detail. As we have seen 
in Sec.~\ref{sec:RN}, for the RN 
black hole, a phenomenon of flux expulsion was observed for small
enough black holes; essentially the event horizon is an infinite proper
distance away, and provided the overall radius of the black hole sits
roughly in the core of the string, it is easier for the magnetic flux
lines of the massless vector field in the core to avoid the black hole than
pierce it -- it is only once the boson becomes massive outside the string
core that the 
energy scales tip the other way. Thus, do we get the same phenomenon
here? There is reason to believe we should. In an elegant construction,
Wald, \cite{Wald}, showed how Killing fields generate probe electromagnetic
fields on Ricci flat backgrounds, and presented a particular solution
which represents a constant axial magnetic field threading the black hole.
The gauge potential has the form
\be
A^\mu \propto \left (2a \partial_t + \partial_\varphi\right)^\mu\,,
\ee
which generates a uniform magnetic field $B_z$ far from the black hole, 
and has zero nett
charge on the black hole. The electric field is not however vanishing, and 
instead sweeps down the axes and out along the equator. While the 
flux lines of the Wald solution cross the horizon for nonextremal black
holes, for {\it all} extremal black holes, the flux is expelled.

Clearly the Wald solution only works for a massless vector field, 
however, one could argue that for small black holes which are well
below the scale of the string, the black hole will be sitting in the string
core, and should therefore see the gauge field as effectively massless
and hence repel it giving rise to flux expulsion. To some extent this
interpretation is correct, however, the situation is a great deal more
complex. In the Wald solution, the photon is massless throughout the
whole of spacetime, whereas for the string, the gauge field is only 
approximately massless inside the string core. Thus the Wald electric
flux, which sweeps down from the poles and out at the equator, now cannot
correspond to an electrically neutral black hole inside the string core.
We therefore cannot simply use the Wald expression as an approximate core
solution. Nonetheless, we find a similar expulsion phenomenon occurs,
although the numerical sensitivity of the low mass black hole system
leads us to suspect that there is a dynamical instability in the small
extremal black hole, analogous to the Kerr-adS instability \cite{CD}.
It is interesting that extremal Kerr is so different from extremal RN,
however, perhaps not surprising due to the rather different structure of
the near horizon spacetime.

\subsection{Near horizon expansion}

To study the near horizon limit, it is useful to rewrite the vector field
in terms of the alternative variables $P$ and $Q$:
\be
\begin{aligned}
P &= P_\varphi + a \sin^2\!\theta P_t\,, \\
Q &= \rho^2 P_t + a P_\varphi\,,
\end{aligned}
\ee
giving
\bea
\Bigl[  \frac{P^2}{\Sigma \sin^2\!\theta}\!\!&-&\!\!
\frac{Q^2}{\Sigma\Delta}\Bigr] X =
\frac{\Delta}{\Sigma} X_{,rr} + \frac{2(r-GM)}{\Sigma} X_{,r}
+\frac{X_{,\theta\theta}}{\Sigma} + \frac{\cot\theta X_{,\theta}}{\Sigma}
+ \frac12 X(1-X^2)\,, \label{XPQ}\qquad\quad\\
\frac{X^2 P}{\beta} &=& \frac{\Delta}{\Sigma} P_{,rr} 
+ \frac{P_{,\theta\theta}}{\Sigma}
-\frac{\cot\theta P_{,\theta}}{\Sigma}  \Bigl( 1 
- \frac{2a^2\sin^2\!\theta}{\Sigma}\Bigr) 
+ \frac{2P_{,r}}{\Sigma^2} \bigl[ \Sigma(r-GM) - r\Delta \bigr]\nonumber\\
&& + \frac{2a  \sin^2\!\theta}{\Sigma^2} \Bigl(rQ_{,r} 
-\cot\theta Q_{,\theta} + aP- Q \Bigr)\,,\\
\frac{X^2 Q}{\beta} &=& \frac{\Delta}{\Sigma} Q_{,rr} 
+ \frac{Q_{,\theta\theta}}{\Sigma}
-\frac{\cot\theta Q_{,\theta}}{\Sigma}  
\Bigl(1 - \frac{4GMr}{\Sigma}\Bigr)\nonumber \\
&&+ \frac{2\Delta}{\Sigma^2} \bigl[ \cot\theta (Q_{,\theta} - a P_{,\theta})
-r(Q_{,r}-aP_{,r})+ Q - aP \bigr]\,.
\eea
In particular, for the extremal Kerr black hole $\Delta=(r-r_+)^2$ 
and so, similar to the RN case,  we expand near the horizon
\bea
X &=& \xi_0(\theta) + (r-r_+) \xi_1(\theta) + \dots\,,\nonumber\\
P &=& \pi_0(\theta) + (r-r_+) \pi_1(\theta) + \dots\,,\\
Q &=& \psi_0(\theta) + (r-r_+) \psi_1(\theta) + \dots\,.\nonumber
\eea
Eq.~(\ref{XPQ}) (or finiteness of energy on horizon) then implies 
that $\psi_0=0$, and the leading order pieces of each equation read
\bea
\xi_0'' + \cot\theta\xi_0' + \frac{r_+^2}{2} (1+\cos^2\!\theta) 
\xi_0 (1-\xi_0^2) -\Bigl[ \frac{\pi_0^2}{\sin^2\!\theta} - \psi_1^2\Bigr] 
\xi_0 &=&0\,,\label{xi0}\\
\pi_0'' -\cot\theta \frac{3\cos^2\!\theta-1}{1+\cos^2\!\theta}\pi_0' 
+ \frac{2\sin^2\!\theta}{1+\cos^2\!\theta} (\psi_1 + \pi_0)
-\frac{r_+^2}{\beta} \xi_0^2 \pi_0 (1+\cos^2\!\theta)&=&0\,,\label{pi0}\\
\psi_1'' + \cot\theta\frac{3-\cos^2\!\theta}{1+\cos^2\!\theta}\psi_1'
-\frac{r_+^2}{\beta} \xi_0^2 \psi_1 (1+\cos^2\!\theta)&=&0\,.\qquad\quad
\label{psi1}
\eea
Note that although the expansion does not in general decouple from 
the bulk (because of the appearance of $\psi_1$) it does form a closed
system in this extremal case.
The constraints on the solutions are that they must be symmetric around 
$\theta=\pi/2$, and $\xi_0=0$, $\pi_0 =1$ at $\theta=0,\pi$.

\subsection{Flux penetration and expulsion}

Let us first show that for large black holes a string will always 
penetrate the black hole horizon. Similar to the extremal RN case, 
we proceed by contradiction. Returning to the full bulk equation 
\eqref{XPQ}, let us assume that flux expulsion occurs, i.e.\ at 
$r_+=a=GM$ we have $X=0$ and $P_\varphi=1$ (with $P_t=-1/2r_+$)
leading to $P = (1+\cos^2\theta)/2$, and hence $Q'(r_+)=-1$ from (\ref{pi0}). 
Therefore near $r_+$ both $\partial_r (\Delta \partial_rX)>0$, 
and $(Q^2/\Delta\Sigma - X^2/2)>0$.
Hence Eq.\ \eqref{XPQ} implies 
\bea
\frac{1}{2}r_+^2\sin^2\!\theta X\!&+&\!\sin\theta \partial_\theta 
(\sin\theta \partial_\theta X)\nonumber\\
\!\!\!\!\!\!\!\!&\leq&\frac{1}{2}r_+^2(1+\cos^2\!\theta)\sin^2\!\theta X
+\sin\theta \partial_\theta (\sin\theta \partial_\theta X)
<XP^2<X\,.\qquad\ 
\eea  
However, this is the same equation \eqref{cond} as discussed in 
Sec.~\ref{sec:RN} and the discussion therein therefore applies. 
Hence we conclude that for any $r_+>\sqrt{8.5}\approx 2.92$ the 
vortex must pierce the extremal Kerr black hole. 

Let us now look more closely at what happens on the horizon. A simple 
inspection of (\ref{psi1}) shows that if $\xi_0 \neq 0$, 
then $\psi_1=0$. Eqs. \eqref{xi0} and \eqref{pi0} now read
\be
\begin{aligned}
\xi_0'' + \cot\theta\xi_0' + \frac{r_+^2}{2} (1+\cos^2\!\theta) 
\xi_0 (1-\xi_0^2) 
-\frac{\pi_0^2\xi_0}{\sin^2\!\theta}&=0\,,\\
\pi_0'' -\cot\theta \frac{3\cos^2\!\theta-1}{1+\cos^2\!\theta}\pi_0' 
+ \frac{2\sin^2\!\theta}{1+\cos^2\!\theta}
\pi_0-\frac{r_+^2}{\beta} \xi_0^2 \pi_0 (1+\cos^2\!\theta)&=0\,,
\end{aligned}
\label{twoEqs}
\ee
and form a pair of equations purely representing data on the horizon, 
decoupled from the bulk. A general analytic discussion of these equations 
is rather involved. However, let us assume, based on energetic 
considerations -- similar to the extremal RN case, that the 
fields $\xi_0$ and $\pi_0$ have only one turning point on the horizon (a fact 
confirmed to a certain extent by a numerical analysis).  
Then the field $\xi_0$ starts from zero at $\theta=0$ and monotonically 
increases to reach its first maximum at $\theta=\pi/2$, 
$\xi_M= \xi_0(\pi/2)<1$, while the value of $\pi_0$ monotonically 
decreases to reach its first minimum  $\pi_m=\pi_0(\pi/2)<1$. 
Since $\pi_0''(\pi/2)>0$, the second equation at $\theta=\pi/2$ 
implies that $r_+^2 \xi_M^2 - 2\beta >0$, i.e., $r_+^2>2\beta$.
Thus, for $r_+<\sqrt{2\beta}$ the 
penetrating solution cannot exist and expulsion must occur.

To show that the flux expulsion is indeed a solution of our near 
horizon equations \eqref{xi0}--\eqref{psi1} we now consider the case when 
$\xi_0 \equiv 0$. Then $\psi_1={const.}$, and
(\ref{pi0}) has the general solution $\pi_0 = \lambda \sin^2\!\theta 
+\gamma \cos\theta - \psi_1$. Applying the boundary conditions, and 
symmetry around $\pi/2$, then yields $\gamma=0$, $\psi_1 = -1$. 
Moreover, the requirement that the field strength invariant 
$F_{\mu\nu}F^{\mu\nu}$ remains finite at $\theta=0$ implies 
$\lambda=-1/2$. Therefore the solution reads $\pi_0=-\frac{1}{2}
\sin^2\!\theta+1$ and $\psi_0=0$. In the original variables 
this corresponds to 
\bea
P_\phi &=& 1\,,\qquad P_t=-\frac{1}{2r_+}\,,
\eea
on the horizon and hence represents a flux-expelled solution. 
Let us remark that if there is a phase transition between the 
flux penetration and expulsion, the value of $\psi_1$ on the 
horizon necessarily suffers from a discontinuity: $\psi_1=0$ for flux 
penetration whereas $\psi_1=-1$ in the case of expulsion.

Our analytic arguments suggest that similar to the RN case there 
exists a critical radius $r_c$, $1.41<r_c<2.92$ for $\beta=1$, 
below which the flux is necessarily expelled. 
Numerical investigations actually indicate $r_c\approx 1.912$. 
The situation seems, however, slightly different to the RN case. 
Whereas we have seen that for the extremal RN black holes the 
transition was continuous (the fields on the horizon vary smoothly 
between the penetrating and the expelling phase), in the case of 
the extremal Kerr black hole we find a very sharp transition. For
example, for $\beta=1$, we find that for $r_+=1.911855$ a piercing 
solution of Eqs. \eqref{twoEqs} exists with $\pi_m\approx 0.57515$ 
and $\xi_M\approx 0.76166$ whereas there is no piercing solution for
$r_+=1.9118525$.  The phase transition would appear to be 
discontinuous. This is backed up by the fact that 
the first derivative of the $Q$-field, $\psi_1$, is discontinuous
between expulsion and penetration. 

Figure \ref{fig:RNK} shows a comparison of the Kerr and RN 
phase transitions for several values of $\beta$. The order parameter
plotted is the maximum value of the Higgs field,  $X_m=\xi_0(\pi/2)$,
attained on the equator. Interestingly, not only is the nature of the
Kerr phase transition different from that of RN, but also the response
to varying $\beta$ is quite different. While both exhibit a 
lowering of critical radius as $\beta$
drops (because of the gauge core becoming thinner), in contrast
to the RN case the Kerr
black hole has higher critical radius as $\beta$ increases. 
While in both instances for smaller $\beta$ the gauge core 
becomes more diffuse, the nature of the equations governing
the gauge fields is different. For Kerr, it is the combination
of $P_\varphi$ and $P_t$, $\pi_0$, that is determined on the 
horizon, and this has two contributions to its effective local 
mass in \eqref{twoEqs}: One geometric, and one coming from
the Higgs field, which must dominate if $\pi_0$ is not to expel.
Once the flux expels, we require $\psi_1=-1$, and hence $\xi_0=0$.
Since the term involving the Higgs field has a factor $r_+^2/\beta$,
it is clear that increasing $\beta$ will increase the critical expulsion
radius.
We believe that this interesting behavior deserves more 
attention in the future. 

\FIGURE{
\includegraphics[scale=0.65]{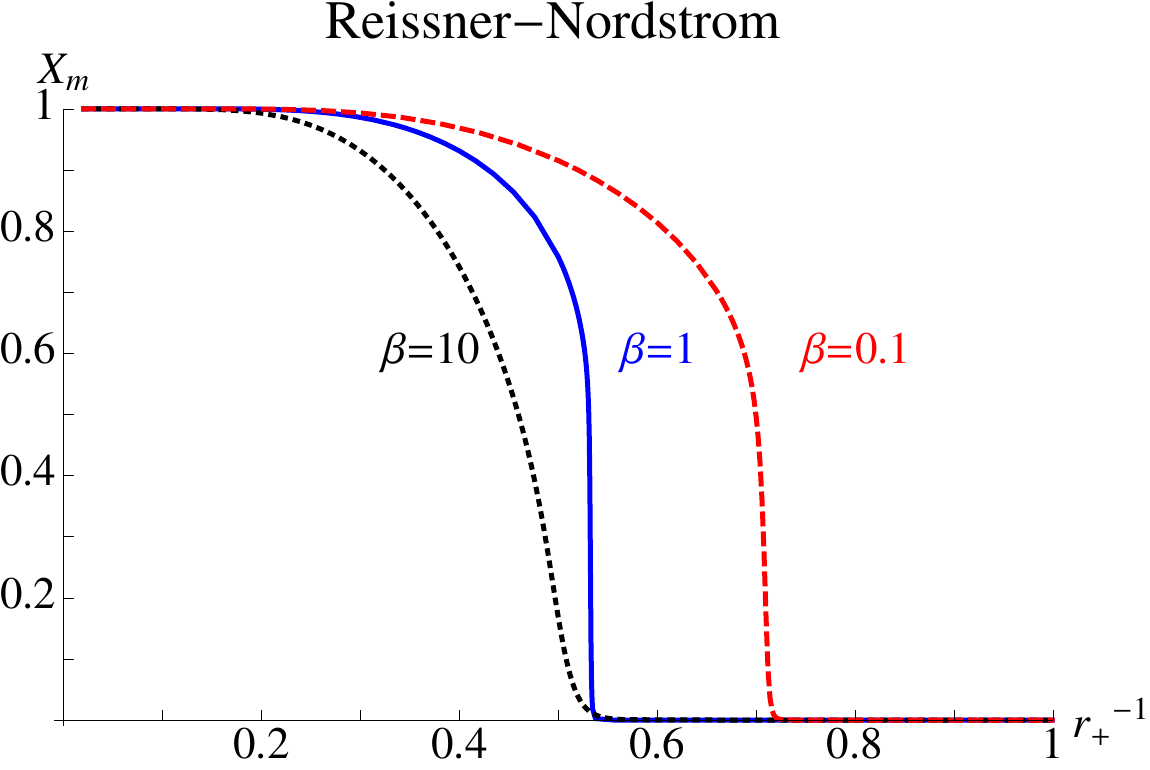}\nobreak
\includegraphics[scale=0.65]{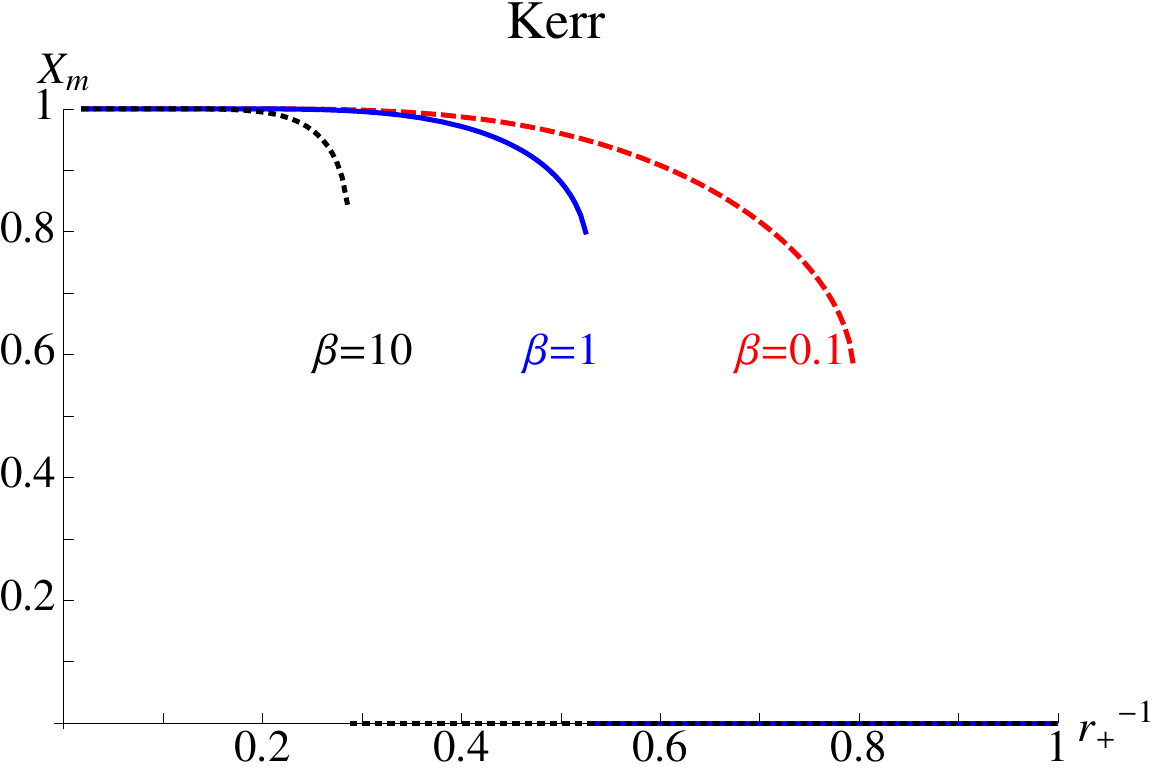}
\caption{Phase plots for the RN and Kerr extremal black holes.
The maximum value of the Higgs field, $X_m= \xi_0(\pi/2)$, is
plotted against the horizon radius $r_+$. The transition is shown
for different values of the Bogomolnyi parameter: $\beta=10$ in
dotted black, $\beta=1$ in solid blue, and $\beta=0.1$ in 
dashed red.}
\label{fig:RNK}
}

\section{Backreaction of the vortex on the black hole}

In the literature, it has been assumed (see e.g.\ \cite{Aliev,GalMas})
that the geometry of a cosmic string threading a Kerr black hole
will simply be given by the Kerr solution with the $\varphi-$angle 
having a reduced range corresponding to the angular deficit.
However, here we will show that this is not in fact the case.
The situation is more subtle, and far more interesting.
In brief, what we show is that the string does indeed induce a
conical deficit, but a deficit from the perspective of an
azimuthal coordinate co-rotating with the black hole, so that the
event horizon of the black hole is a 2-sphere with a wedge removed.
Because the horizon is rotating relative to an asymptotic observer,
this is {\it not} equivalent to a simple angular deficit in the full
spacetime, but rather, there is a more complex response in the region
of the black hole, with the asymptotic deficit angle behaviour recovered
only at large $r$. As a consequence, the ergosphere is shifted, and nearby
orbits of Kerr black holes will also be affected.

Our approach is to use the perturbative technique of \cite{AGK,RGMH},
in which we solve the full Einstein-abelian Higgs system order by
order in $\epsilon = 8\pi G\eta^2$. Strictly, as we wish to present
simple `analytic' expressions, we also solve for a ``thin'' string,
in which $\sqrt{\lambda}\eta r_+ \gg 1$. For small $\epsilon$,
we can use the probe vortex solution to compute the leading order
gravitational backreaction, and for large black holes, we have 
demonstrated that our `analytic' approximation \eqref{ApproxSol}
is an excellent expression which closely mimics the full numerical
solution. In particular, it is a reliable expression both within the core
of the string, and on the horizon of the black hole. 

In order to proceed, we need to express the metric in a useful set 
of coordinates which reflect the axial symmetry of the Kerr-cosmic 
string set-up. For this purpose we shall use the Weyl form of the metric
(see e.g.\ \cite{CLSZ})
\be
ds^2=e^{2\lambda} dt^2-\alpha^2 e^{-2\lambda}\bigl[d\varphi+B dt\bigr]^2
-e^{2(\nu-\lambda)}(dx^2+dy^2)\,,
\ee
where the functions $\alpha, B, \nu$ and $\lambda$ are functions 
of the $x$ and $y$ coordinates only. Note the similarity 
with \eqref{selfgravG}, only there the functions depended only on one
coordinate. 
The Ricci tensor of this metric is given by:
\bea
R^\varphi_\varphi+R^t_t &=& e^{2(\lambda-\nu)}
\frac{\nabla^2 \alpha}{\alpha}\,, \label{alpha}\\
R^t_\varphi &=& \frac{\alpha}{2}e^{-2(\lambda+\nu)}\Bigl[
-3\nabla\alpha\cdot \nabla B+4\alpha\nabla B\cdot \nabla\lambda
-\alpha\nabla^2 B\Bigr]\,,\label{tphi}\\
R_{\varphi\varphi} &=& -\frac{\alpha}{2}e^{-2\nu}
\Bigl[\alpha^3 e^{-4\lambda} (\nabla B)^2
+2\nabla^2\alpha-2\nabla\alpha\cdot \nabla \lambda
-2\alpha\nabla^2\lambda\Bigr]\,,\label{phiphi}\\
R^x_x+R^y_y &=& \frac{e^{2(\lambda-\nu)}}{2\alpha}
\Bigl[2{\nabla^2\alpha}-{\alpha^3e^{-4\lambda}}(\nabla B)^2
+ 4e^{\lambda} \nabla\cdot\left(\alpha \nabla e^{-\lambda}\right)
+4\alpha\nabla^2\nu\Bigr]\,,\label{alpha2}\\
R_{xy} &=& \frac{1}{2\alpha}\bigl[\alpha^3e^{-4\lambda} B_x B_y
-4\alpha\lambda_x\lambda_y+2(\alpha_x\nu_y+\alpha_y\nu_x)
- 2\alpha_{xy}\Bigr]\,,\label{xy}\\
R_y^y &=& \frac{e^{2(\lambda-\nu)}}{2\alpha}\Bigl[-
\alpha^3e^{-4\lambda} B_y^2
+2\alpha\nabla^2(\nu-\lambda)-2(\alpha_y\nu_y-\alpha_x\nu_x)\nonumber\\
&&\qquad-2\nabla\alpha\cdot \nabla\lambda+4\alpha\lambda_y^2
+2\alpha_{yy}\Bigr]\,,\label{yy}
\eea 
where we have introduced the two-dimensional gradient 
operator $\nabla=(\partial_x, \partial_y)$ as well as the 
corresponding dot product, expressing for example the Laplace 
operator as $\nabla^2=\nabla\cdot \nabla=\partial_x^2+\partial_y^2$. 

In particular, defining\footnote{Note this is not the usual Weyl gauge,
in which the $\alpha$ variable is typically equal to one of $x$ or $y$,
however, this choice proves easier to analyse, and is closer to the
standard Boyer Lindquist Kerr gauge.}
\be
x=\int \frac{dr}{\sqrt{\Delta}}\,,\quad y=\theta\,,
\ee 
the background (Kerr) solution can be written as
\be
%\begin{aligned}
\alpha_0 =\sqrt{\Delta}\sin\theta\,,\quad  B_0
=-\frac{2aGMr}{\Gamma}\,,\quad
e^{2\nu_0} =\frac{\Delta \Sigma^2}{\Gamma}\,,
\quad e^{2\lambda_0}=\frac{\Delta \Sigma}{\Gamma}\,.
%\end{aligned}
\label{WeylKerr}
\ee
The procedure for finding the back-reacted vortex solution is to 
use the analytic approximation \eqref{ApproxSol} to find the energy
momentum of the vortex solution, which will be a good approximation to
the true energy momentum, and to use this to find the leading order
correction to the metric by expanding the Einstein equations, 
\be
R_{\mu\nu}=\epsilon \Bigl(T_{\mu\nu}-\frac{1}{2}Tg_{\mu\nu}\Bigr)\,,
\quad \epsilon\equiv 8\pi G_N \eta^2\,,
\ee
around 
the background Kerr solution using the Weyl expressions \eqref{WeylKerr}.

In this limit, the energy momentum tensor is found to leading order\footnote{To
derive these forms, we have computed the components of $T_{\mu\nu}$ using
the analytic approximation, and have expanded the metric coefficients
to leading order near the string core, so that for example $\Sigma =
r^2+a^2\cos^2\theta = \rho^2 - a^2 R^2 / \rho^2 = \rho^2(1+{\cal O}
(r_+^{-2}))$.} to be
\be
\begin{aligned}
T^t_t&\approx T^x_x\approx X_0'^2
+\frac{X_0^2P_0^2}{R^2}+\beta\frac{{P_0'}^2}{R^2}
+\frac{1}{4}(X_0^2-1)^2\equiv {\cal E}\,,\\
T^y_y&\approx -X_0'^2+\frac{X_0^2P_0^2}{R^2}
-\beta\frac{P_0'^2}{R^2}+\frac{1}{4}(X_0^2-1)^2 \equiv -{\cal P}_R \,,\\
T^\varphi_\varphi&\approx X_0'^2-\frac{X_0^2P_0^2}{R^2}
-\beta\frac{P_0'^2}{R^2}+\frac{1}{4}(X_0^2-1)^2\equiv -{\cal P}_\phi \,,\\ 
T_{xy} &\approx \frac{\sqrt{\Delta}r}{\rho} R ({\cal E} + {\cal P}_R)\,,
\\
T^t_\varphi&\approx - \frac{4GMra}{\rho^8} \left [ (\rho^2 - 4r^2 ) RPP'
- a^2 R^2 (X^2P^2 + P'^2) \right ]\approx 0\,,
\end{aligned}
\label{KerrTab}
\ee
where ${\cal E}$ etc.\ denote the energy-momentum components of the 
Nielsen-Olesen vortex, defined in \eqref{NOTab}, which are simply functions
of $R=\rho\sin\theta$. 
Because these components are functions of the $R$ variable only, this
leads to a modification of the Kerr geometry which is also dependent on
$R$. 

In order to motivate the form of the perturbed metric, we first 
note that at large ``$r\cos\theta$'', the metric should approach 
the form of the isolated gravitating vortex, given in 
(\ref{SGalpha},\ref{SGlambda}). Thus we expect 
that $\delta\alpha = \epsilon\alpha_0 \alpha_1(R)$,
$\delta\lambda = \delta\nu/2=\epsilon\lambda_1(R)$. 
Of course we must confirm that the equations of motion indeed
lead to perturbations of this form. 

First, consider the Einstein equation \eqref{alpha},
\be
\delta(R^t_t+R^\varphi_\varphi) = \frac{e^{2(\lambda-\nu)}}{\alpha }
\nabla^2 \delta \alpha = \frac{\nabla^2 \delta\alpha}{\sqrt{\Delta}\rho R}
= - \epsilon \, ({\cal E}-{\cal P}_R)\,,
\ee
which is solved to leading order by a perturbation of the form
\be
\alpha=\alpha_0\Bigl(1+\epsilon \alpha_1(R)+O(\epsilon^2)\Bigr)\,.
\ee
where $\alpha_1$ satisfies
\be
\alpha_1'' +2\frac{\alpha_1'}{R}= - ({\cal  E}-{\cal P}_R) \;\;\; 
\Rightarrow \;\;\; \alpha_1 = - \int_0^R \!\!\! R({\cal E}-{\cal P}_R) dR 
+ \frac1R \int_0^R \!\!\! R^2 ({\cal E}-{\cal P}_R) dR\,, 
\ee
which is in fact identical in form to the self-gravitating correction
\eqref{SGalpha}.

Next, noting that $ B$ and its derivatives are subdominant, we obtain
for the next Einstein equations \eqref{phiphi}, \eqref{alpha2} and \eqref{yy}
\bea
\delta R_{\varphi\varphi} &=& -\epsilon R\left [ R\alpha_1'' +
2 \alpha_1' - R\lambda_1''- \lambda_1' \right] = R^2 \left[
{\cal E} + \frac12({\cal P}_\phi - {\cal P}_R ) \right ]\,,\\
\delta(R^x_x+R^y_y) &=& \epsilon \left [ \alpha_1'' + \frac{\alpha_1'}{R} 
+ 2 \nu_1'' - 2 \lambda_1'' - 2 \frac{\lambda_1'}{R} \right ] 
= -\epsilon \left [ {\cal E} - {\cal P}_\phi \right ]\,, \\
\delta R^y_y &=& \epsilon \left [ \alpha_1'' + \frac{\alpha_1'}{R} 
+ \nu_1'' - \lambda_1'' - \frac{\nu_1'}{R} - \frac{\lambda_1'}{R} \right ] 
= -\epsilon \left [ {\cal E} - \frac12({\cal P}_\phi - {\cal P}_R ) \right ]\,,\qquad
\eea 
the first of which gives
\be
\lambda_1 = \frac12 \int_0^R R{\cal P}_R dR\,,
\ee
and consistency with the next two implies $\nu_1=2\lambda_1$, also
as in the self-gravitating case.

Now we can examine the variation $\delta B$, which cannot be deduced 
from an asymptotic analysis, since the background function 
$ B={\cal O}(r^{-3})$ is subdominant. 
Instead, we must examine the near horizon behaviour of \eqref{xy}, 
in which the only term not explicitly convergent is
\be
\alpha_0^2 e^{-4\lambda_0}  B_{0,x} \delta B_{,y} 
\sim -\frac{2GMa}{\rho^2\sqrt{\Delta}} \left ( 1 - \frac{4r^2}{\rho^2}\right)
R^2 \delta B_{,y}\,.
\ee
Given the RHS of this equation from \eqref{KerrTab},
we quickly see that we cannot find a form of $\delta  B$ which has
the requisite functional dependence on the background coordinates, as well
as on the variable $R$. In particular, it is transparent that if we set
$\delta B = \epsilon  B_0 \alpha_1^{-1}$, which is what would be
required for a pure $\varphi$-angular conical deficit, then this
would lead to a divergence in $\delta R_{xy}$ at the horizon, and would
not solve the Einstein equations.
Thus $\delta B=0$. This simple result will
have significant consequences as we will see.

We can now check the remaining equations:
\bea
\delta R_{xy} &=& -\epsilon\sqrt{\Delta} \;\frac{r}{\rho}
\left[ R\alpha_1'' + 2\alpha_1' - 4 \lambda_1' \right]
= \epsilon\sqrt{\Delta} \;\frac{r}{\rho}
\left [ {\cal E} + {\cal P}_R \right ]\,, \\
\delta R^t_\varphi &=& \epsilon \frac{GMra}{\rho^8} \left ( \rho^2 - 4r^2 
+ 2a^2 \right ) R^3 \left [ 3\alpha_1'-4\lambda_1' \right]\\
&=& - \epsilon \frac{4GMra}{\rho^8} \left [ (\rho^2 - 4r^2 ) RPP'
- a^2 R^2 (X^2P^2 + P'^2) \right ] ={\cal O} (r_+^2/r^5)\nonumber\,.
\eea 

Pulling all the details together, and looking outside the core of the vortex,
we see the asymptotic form of the Kerr-vortex is
\be
\begin{aligned}
ds^2 =& \left ( 1 - \frac{2GMr}{\Sigma} + \frac{8 (GMar\sin\theta)^2}
{\Gamma\Sigma} \epsilon {\hat \mu} \right ) dt^2
-\!\Sigma d\theta^2
-\!\frac{\Sigma}{\Delta}dr^2\\
&-\frac{\Gamma}{\Sigma}(1-2\epsilon {\hat \mu} )\sin^2\!\theta \,d\varphi^2\!
+\frac{4GMar\sin^2\!\theta}{\Sigma} (1-2\epsilon {\hat \mu}) dtd\varphi\,,
\end{aligned}
\ee
where $\hat \mu$ is the renormalised energy per unit length of the 
string defined in \eqref{muhat}.
It is clear that while there is an angular deficit in this spacetime, which
does approach the standard conical deficit at large distances,
in the vicinity of the black hole, as far as these Boyer-Lindquist
coordinates are concerned, the deficit is felt not only by the `angular'
$\varphi$-coordinate, but also by the time component of the metric.
While this seems a little strange and worrying, if we instead transform
to a frame co-rotating with the black hole, $\varphi_H = \varphi - 
\Omega_H t = \varphi -  B(r_+)$, then the effect of the cosmic string
is indeed to remove a deficit angle -- {\it from the perspective of the
black hole}. Note this is not in contradiction
with the Schwarzschild result, it is simply that in Schwarzschild there
is no difference between the spatial angular variable on the horizon
and that at infinity.

\section{Discussion}

To sum up: we have shown how to correctly thread a rotating black
hole with vortex hair. A consequence of rotation is that the angular
and time components of not only the metric, but also the vortex
fields are interconnected. This leads to an electric field in the polar
regions of the black hole. That this is a genuine electric flux, and
not some frame dragging transformation effect is easily verified by
computing 
\be
|F\wedge F| \sim {\bf E} \cdot {\bf B}\sim 
\frac{8GMaP_0(R)P_0'(R)(3r^2-a^2)}{R\rho^6}
\ee
for the approximate solution -- clearly a nonvanishing quantity.
(We have also checked ${\bf E}\cdot{\bf B}$ for the numerical
solution, but as can be anticipated from the excellent agreement
between numerical and approximate solutions, this gives 
roughly the same result.) Thus, as with the Wald solution,
there is clearly an induced electric field in all frames.

We also explored the flux expulsion transition for the Kerr black 
hole, and while we observed flux expulsion, the gradient flow
method became extremely sensitive, particularly around the
phase transition, and took several orders of magnitude longer
in `time' to converge, as well as requiring several orders of magnitude
smaller `time' steps in the program. This is in contrast to the
extremal RN solution, which, while being a little more sensitive to
find numerically, is more or less in the same ball park of convergence
and sensitivity as the Schwarzschild case. We conjecture that this
is due to a super-radiant instability of the rotating black hole within the 
vortex core. Kerr-adS black holes exhibit an instability, \cite{CD}, 
due to the confining nature of the adS spacetime. Here, we do not
have a negative cosmological constant, however, we do have 
confinement: exterior to the vortex core the scalar and gauge fields
are massive, so any perturbation will primarily propagate up and down
the string, as a massless zero mode. Modes transverse to the
string will be reflected back. Thus, we can envisage a range of modes
which get reflected back from the vortex edge back to the rotating Kerr
black hole, picking up more angular momentum, getting reflected back
and so on. It would be interesting to explore this suspicion.
\FIGURE{
\includegraphics[scale=0.7]{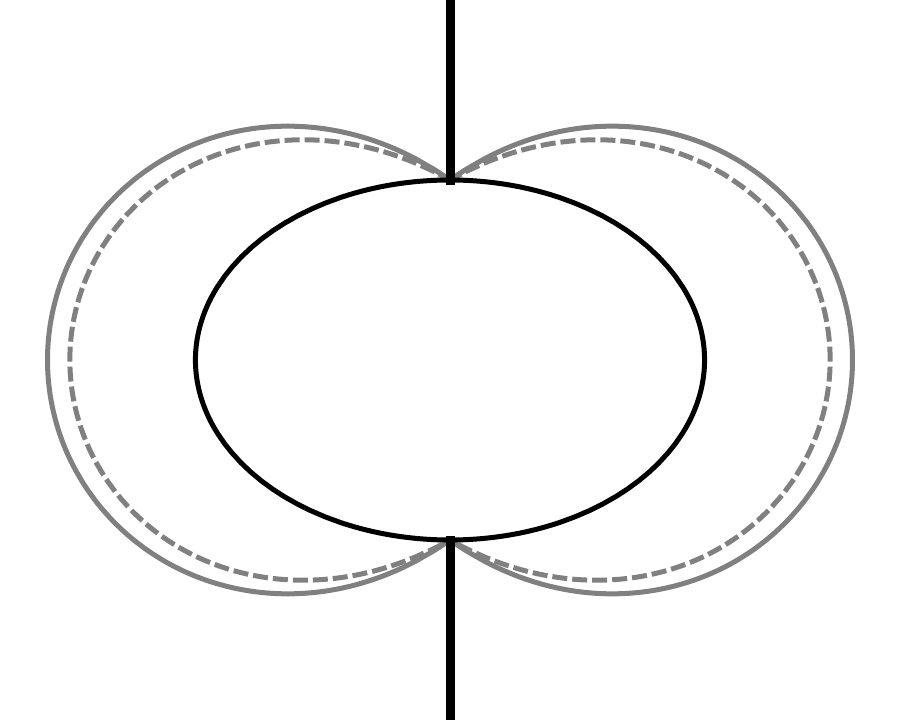}
\caption{Shifting of the ergosphere -- depicted here for an extremal 
Kerr black hole with $\epsilon=0.2$.}
\label{fig:ergo}
}

Perhaps the most important result from our study however is
the discovery that the conical nature of the spacetime is not in
the form of a simple deficit angle at infinity. All previous studies
have assumed that the string will simply redefine the nature of the
$\varphi$-angle, but we find this is not the case. Of course, once
spotted, this seems entirely natural, due to frame dragging around
the black hole, but we stress that this is a new physical phenomenon
and will lead to distinct features of the  Kerr+vortex black hole.
For example, the ergosphere will be shifted (see figure \ref{fig:ergo})
and orbits around the black hole will be affected\footnote{Note, this
is a distinct effect from that considered recently in \cite{GMP},
in which cross terms in the metric are introduced via
generating transformations.}. We have plotted
in figure \ref{fig:isco} the impact of the correction on the ISCO's of the 
Kerr black hole. However, the
minute value of the cosmic string tension will most likely render
this effect outside the range of observational precision.
\FIGURE{
\includegraphics[scale=0.7]{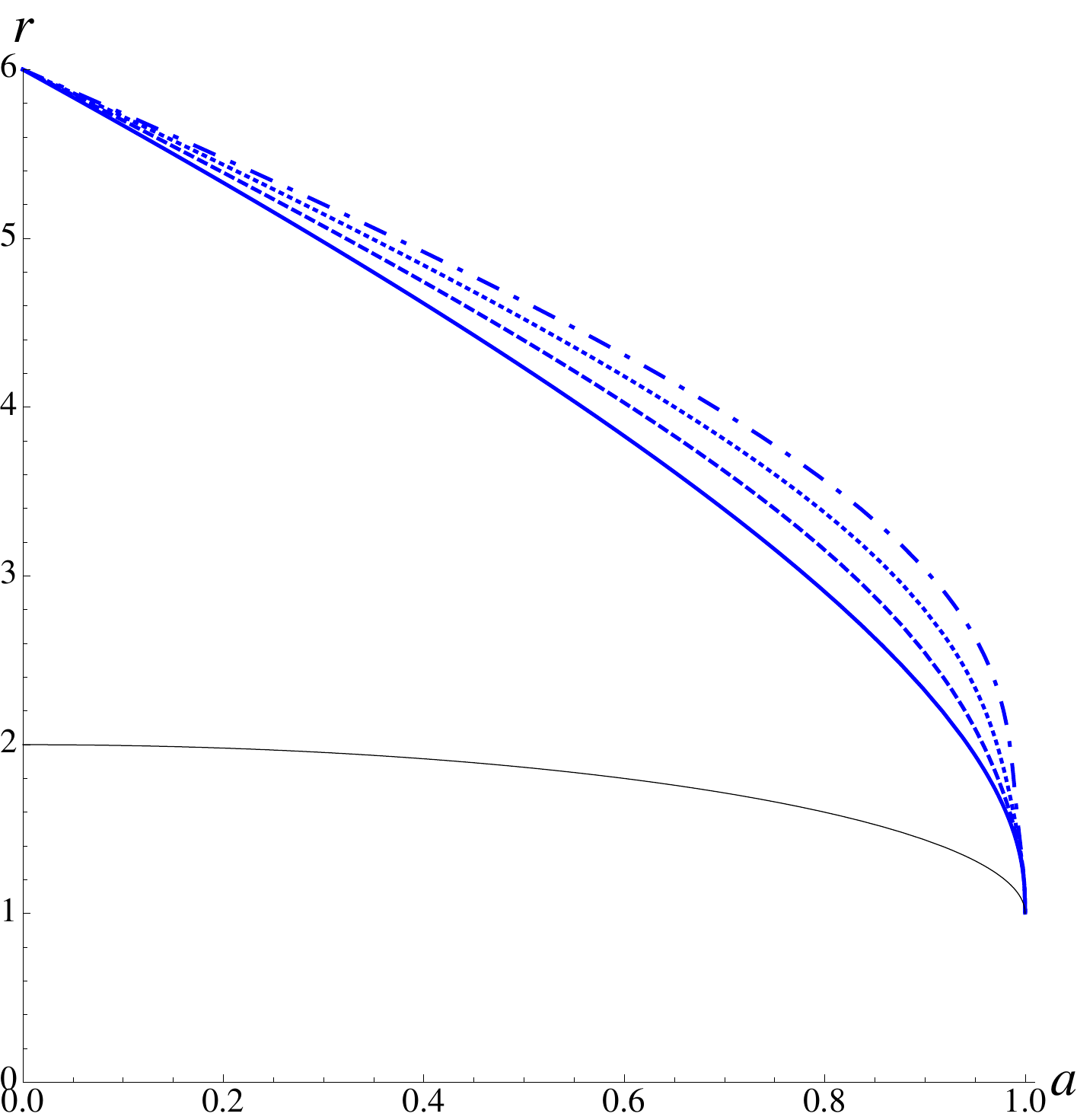}
\caption{Shifting of ICSO's as a function of $a$ (measured in 
units of $GM$) for $\epsilon=0, 0.1, 0.2, 0.3$, represented by
solid, dashed, dotted, and dot-dashed blue lines respectively.
The black line indicates the location of the event horizon.}
\label{fig:isco}
}

All in all, the vortex-Kerr system has proved to be surprisingly
different and much more interesting that the standard Schwarzschild
black hole hair.

\acknowledgments
We would like to thank Avery Broderick, Roberto Emparan and 
Paul Sutcliffe for helpful discussions.
RG is supported in part by STFC (Consolidated Grant ST/J000426/1),
in part by the Wolfson Foundation and Royal Society, and in part
by Perimeter Institute for Theoretical Physics. DK is supported by 
Perimeter Institute. DW is supported by an STFC studentship. 
Research at Perimeter Institute is supported by the Government of
Canada through Industry Canada and by the Province of Ontario through the
Ministry of Economic Development and Innovation.

\providecommand{\href}[2]{#2}\begingroup\raggedright\endgroup

\end{document}